\documentclass[12pt]{article}
\usepackage[margin=1in]{geometry}
\usepackage{times}
\usepackage{natbib}
\usepackage{etoolbox}
\usepackage[
  colorlinks=true,
  linkcolor=blue,
  citecolor=blue,
  urlcolor=blue
]{hyperref}
\usepackage{url}
\bibpunct{(}{)}{;}{a}{,}{,}
\hypersetup{
  colorlinks=true,
  citecolor=blue,
  linkcolor=blue,
  urlcolor=blue
}

\makeatletter
\let\prooforiginalNATparse\NAT@parse
\renewcommand{\NAT@parse}[1]{%
  \prooforiginalNATparse{#1}%
  \patchcmd{\NAT@name}{ and }{ \& }{}{}%
}
\makeatother

\title{LLMs for qualitative and mixed-methods\\ social network analysis}
\author{Moses Boudourides \\
School of Professional Studies, Northwestern University,\\
405 Church St., Evanston, IL 60201, USA.\\
\url{Moses.Boudourides@northwestern.edu}}
\date{}

\begin{document}

\maketitle

\begin{abstract}
This manuscript explores the integration of Large Language Models (LLMs) into the field of qualitative and mixed-methods social network analysis (SNA). We argue that the primary focus of this integration should be on enhancing the depth and rigor of qualitative SNA, rather than on replacing human researchers with automated systems. We begin by outlining the core principles of qualitative and mixed-methods SNA, emphasizing the importance of understanding the meaning of ties, the role of narratives, and the significance of relational identities. We then discuss how LLMs can be used as powerful tools to augment this work, from assisting with data collection and coding to supporting theory-building and abductive reasoning. We also address the limitations and ethical challenges of using LLMs in this context, including issues of bias, hallucination, and the need for reflexivity. We conclude with a series of research designs and practical recommendations for researchers who want to integrate LLMs into their work in a thoughtful and responsible way.
\end{abstract}


\section{Motivation and scope}

\subsection{Why LLMs matter for social network analysis now}

The confluence of large-scale data availability and advanced computational methods has created a new frontier for social network analysis (SNA). While quantitative approaches have long dominated the field, offering powerful tools for understanding network structures, they often fall short of capturing the rich, nuanced meanings embedded in social relationships \citep{borgatti2009network}. The recent advent of Large Language Models (LLMs) presents a pivotal opportunity to bridge this gap. LLMs, with their unprecedented ability to process and generate human-like text, offer a means to analyze the qualitative dimensions of social networks at a scale previously unimaginable. This is not merely a matter of automating existing qualitative methods, but of fundamentally rethinking how we approach the study of social networks. As we will argue, the integration of LLMs into SNA is not a replacement for rigorous qualitative inquiry, but a powerful augmentation that allows us to ask new questions and uncover deeper insights into the complex interplay of social structure and meaning \citep{abbott2004methods,salganik2018bit}. Throughout this article, statements about LLM capabilities refer to the state of publicly available models at the time of writing and should not be interpreted as permanent properties of language models.

\subsection{From ties to meaning: Why social networks need qualitative augmentation}

A social network is more than a collection of nodes and edges \citep{borgatti2018analyzing}. It is a dynamic system of relationships, imbued with cultural, social, and personal significance. Qualitative methods have long been essential for uncovering this ``thick description'' of social life \citep{lincoln1985naturalistic,maxwell2012qualitative}, providing context and depth to the structural patterns revealed by quantitative SNA. Ethnography, in-depth interviews, and narrative analysis allow researchers to understand the subjective experiences of actors within a network, the cultural norms that shape their interactions, and the stories they tell about their relationships. Without this qualitative dimension, network analysis risks becoming a sterile exercise in structural formalism, detached from the lived realities of the social world. The challenge, however, has always been one of scale. The time-intensive nature of qualitative research has traditionally limited its application to small, bounded networks. This is where the potential of LLMs becomes most apparent: they offer a means to bring the richness of qualitative inquiry to the scale of large networks, allowing us to analyze the content of communications, the nuances of language, and the emergent properties of meaning across thousands or even millions of interactions \citep{salganik2018bit,small2011lost}.

\subsection{What LLMs do \emph{and do not} replace in SNA}

It is crucial to approach the integration of LLMs into SNA with a clear understanding of their methodological role and current limitations. Rather than attempting to define the absolute limits of LLM capabilities -- which evolve rapidly with each model generation -- we argue that the limits of their application should be defined by the requirements of qualitative methodology itself. Even if future models achieve human parity in contextual reasoning, qualitative methodology requires these responsibilities to remain with the human researcher. They can automate the more laborious aspects of qualitative data analysis, such as coding and memoing, but they cannot substitute for the researcher's theoretical sensitivity, interpretive judgment, and ethical reflexivity. This collaborative model, we argue, is the most promising path forward for a truly mixed-methods approach to social network analysis \citep{abbott2004methods,nelson2020computational,salganik2018bit,small2011lost}.

\section{Qualitative and mixed-methods SNA}

\subsection{Classic qualitative SNA traditions}

Qualitative social network analysis is not a new field \citep{hollstein2011qualitative}. It has a rich intellectual history, with roots in the ethnographic and anthropological studies of the early 20th century. Researchers in this tradition have long recognized that social structures are not merely patterns of connection, but are actively produced and reproduced through meaningful social interaction. From the pioneering work of the Manchester School in the mid-20th century, which emphasized the role of conflict and power in shaping social networks, to the more recent development of relational sociology \citep{emirbayer1997manifesto}, the common thread has been a commitment to understanding the endogenous, processual nature of social life. This tradition stands in contrast to the more structuralist and quantitative approaches that have often dominated the field of SNA, offering a much-needed corrective to the reification of network structures and a deeper appreciation for the role of human agency in shaping social life \citep{white2008identity,crossley2011towards}.

\subsection{Ethnographic networks and meaning of ties}

Ethnography, with its emphasis on long-term immersion in a social setting, is the quintessential method for understanding the meaning of ties in a social network \citep{mitchell1969networks}. By observing and participating in the daily lives of individuals, ethnographers can gain a deep understanding of the cultural norms, social practices, and personal histories that give shape and substance to social relationships. This is not simply a matter of asking people about their networks; it is about observing how those networks are enacted in practice, how they are negotiated and contested, and how they change over time. The ethnographic approach allows us to move beyond the simplistic binary of tie/no-tie to a more nuanced understanding of the qualitative dimensions of relationships, including their strength, multiplexity, and affective content. It is through this deep, contextualized understanding that we can begin to see how the micro-level interactions of individuals give rise to the macro-level structures of the social world \citep{small2011lost}.

\subsection{Narratives, roles, positions, and relational identities}

Social networks are not just structures of interaction; they are also structures of meaning. The stories we tell about our relationships, the social roles we inhabit, and the relational identities we construct are all crucial components of the social fabric \citep{pachucki2010cultural}. Harrison White's relational sociology, particularly his later work on identity and control, provides a theoretical framework for understanding how identities and meanings emerge within networked social processes \citep{white2008identity}. White argues that identities are not fixed attributes of individuals, but are emergent properties of social networks, forged in the ongoing process of social interaction. By analyzing the narratives that people use to make sense of their social worlds, we can gain insight into the cultural and social logics that underpin network structures. This focus on narrative and identity is a hallmark of qualitative SNA, and it is an area where LLMs have the potential to make a significant contribution by allowing us to analyze large corpora of text and identify the recurring themes, narrative patterns, and relational identities that emerge from the data.

A crucial distinction in qualitative social network analysis is between the \textit{meaning} of a tie and its \textit{measurement}. Qualitative inquiry often reveals that what counts as a relationship, how it is recognized, and what it signifies varies across actors and contexts. These interpretive differences directly affect how ties are operationalized, aggregated, or even whether they are represented as ties at all. Making this distinction explicit highlights that network measures are not neutral descriptors but the result of prior interpretive decisions grounded in qualitative understanding \citep{abbott2004methods}.

\subsection{Mixed-methods designs in network research}

\subsubsection{Design logics and integration strategies}

The integration of qualitative and quantitative methods has become increasingly common in social network analysis, as researchers have come to recognize the limitations of relying on a single methodological approach \citep{bellotti2015qualitative,edwards2010mixed}. Mixed-methods designs allow researchers to combine the strengths of both qualitative and quantitative methods, using qualitative data to provide context and depth to quantitative findings, and quantitative data to provide a broader, more generalizable picture of the social landscape. There are a variety of mixed-methods designs that can be used in network research, including sequential designs, in which one method is used to inform the other, and parallel designs, in which both methods are used concurrently \citep{creswell2014mixed}. The choice of design will depend on the specific research questions and the nature of the data being collected. The key, however, is to ensure that the two methods are not simply used in parallel, but are fully integrated, so that the insights from one method can inform and enrich the insights from the other.

\subsubsection{Ego-centered and comparative qualitative network designs}

A particularly important analytic strategy in qualitative and mixed-methods network research is the use of ego-centered networks, where individual actors and their immediate relational environments are treated as cases \citep{crossley2015egonets,boissevain1974friends}. In this approach, egos are not merely nodes embedded in a larger structure but analytic units through which relational meanings, practices, and constraints are examined in depth. Ego-centered qualitative networks allow researchers to combine detailed interpretive analysis with systematic comparison across cases, making them especially well-suited for mixed-methods designs that bridge narrative data, interviews, and network measures.

Qualitative and mixed-methods SNA also rely on systematic comparison across networks, particularly in small-N or case-based research designs. Rather than seeking statistical generalization, this approach compares relational configurations, tie meanings, and positional patterns across a limited set of networks or ego-centered cases. Such comparative network analysis supports analytic generalization by identifying recurring relational mechanisms, contrasts, and boundary conditions across cases, while preserving the contextual richness of qualitative inquiry \citep{crossley2011towards,small2011lost}.

\subsubsection{Interpretation, validation, and reflexivity in network analysis}

An additional practice in qualitative and mixed-methods network research is the use of member validation or respondent feedback in the interpretation of network data. Researchers may present network representations, summaries, or interpretations to participants in order to assess whether these resonate with their lived experiences and relational understandings. Such feedback can reveal mismatches between analytic representations and emic perspectives, prompting revisions to tie definitions, network boundaries, or interpretive claims. Member validation thus functions as an important check on interpretive validity in network analysis \citep{small2011lost}.

Qualitative and mixed-methods SNA also require an explicit distinction between emic and etic perspectives in network construction and interpretation. Emic perspectives capture how actors themselves understand, describe, and enact their relationships, while etic perspectives reflect the analytical categories and representations imposed by the researcher. Making this distinction explicit helps clarify where network definitions, tie categories, and measures align with participants' understandings and where they reflect analytic abstraction. Attending to emic--etic tensions is particularly important when integrating qualitative insight with formal network models \citep{crossley2011towards}.

In qualitative and mixed-methods SNA, network visualizations function not only as descriptive summaries but also as interpretive devices \citep{freeman2000visualizing}. Visual representations can support sense-making by highlighting relational patterns, positions, and anomalies that prompt further qualitative inquiry. Researchers often use visualizations iteratively, adjusting layouts, thresholds, or categorizations in response to emerging interpretations. Treated in this way, network visualizations become part of the analytic process rather than merely tools for presentation \citep{salganik2018bit}.

Concepts such as roles, positions, and block models should also be understood as interpretive constructs in qualitative and mixed-methods SNA, rather than as purely formal outputs. While these techniques identify regularities in relational structure, their substantive meaning depends on contextual and qualitative interpretation. Researchers often draw on interviews, narratives, or observational data to interpret what particular roles or positions signify within a given social setting. Treating these constructs as interpretive bridges helps integrate formal network analysis with qualitative understanding \citep{white2008identity,crossley2011towards}.

Qualitative and mixed-methods network research often involves iterative feedback loops in which qualitative findings lead to revisions of network boundaries, tie definitions, or analytic focus. As researchers gain deeper interpretive insight into how relationships are enacted and understood, they may redefine what counts as a relevant actor, relationship, or interaction. These revisions can alter the structure of the network under analysis and, in turn, prompt further qualitative inquiry. Such feedback loops highlight the dynamic and reflexive nature of network construction in qualitative research \citep{abbott2004methods}.

Assessing validity in qualitative and mixed-methods network research requires attention to forms of validity that are specific to relational data. Beyond statistical or measurement validity, researchers must consider interpretive validity (whether network representations capture participants' meanings), relational validity (whether ties and positions adequately reflect social relations), and integrative validity (whether qualitative and quantitative components are coherently aligned). Making these validity dimensions explicit strengthens the methodological rigor of mixed-methods SNA \citep{creswell2014mixed,small2011lost}.

\subsection{Where qualitative SNA breaks at scale}

Despite its many strengths, qualitative SNA has always faced the challenge of scale. The time-intensive nature of ethnographic fieldwork, in-depth interviewing, and narrative analysis has traditionally limited the application of these methods to small, bounded networks. While this has allowed for a deep and nuanced understanding of these specific cases, it has made it difficult to generalize the findings to larger populations or to compare across different social contexts. This is where the promise of LLMs is most apparent. By automating some of the more laborious aspects of qualitative data analysis, LLMs have the potential to dramatically increase the scale at which qualitative SNA can be conducted. LLMs thus offer a powerful tool for extending the reach of qualitative inquiry, allowing us to ask new questions and explore new frontiers in the study of social networks \citep{salganik2018bit}.

\section{What LLMs add (and where they fail)}

\subsection{LLMs as relational text analyzers}

At their core, LLMs are powerful engines for text-based relational analysis, rather than relational modeling. They can be used to identify not just entities and concepts within a text, but also the relationships between them. This is a crucial capability for SNA, as it allows us to move beyond simply counting co-occurrences to a more nuanced understanding of the relational dynamics at play. For example, an LLM could be trained to distinguish between different types of social ties (e.g., friendship, kinship, professional collaboration) based on the linguistic cues in a text. It could also be used to identify the sentiment or affective tone of a relationship, or to track how relationships change over time. This ability to extract relational information from unstructured text opens up a wide range of new possibilities for SNA, allowing us to analyze large-scale datasets that were previously inaccessible to qualitative researchers \citep{nelson2020computational,salganik2018bit}.

\subsection{Entity extraction vs. relational interpretation}

While LLMs are adept at entity extraction, it is important to distinguish this from the more complex task of relational interpretation. Entity extraction is the process of identifying named entities in a text, such as people, organizations, and locations. Relational interpretation, on the other hand, is the process of understanding the meaning and significance of the relationships between those entities. This is a much more challenging task, as it requires a deep understanding of the social and cultural context in which the relationships are embedded. An LLM might be able to identify that two people are connected, but its ability to determine the full meaning of that connection remains model-dependent and contingent on the explicitness of the textual cues available. Questions such as whether a relationship is one of power and subordination, love and intimacy, or rivalry and competition have become more tractable for current large-scale models, which can classify relational valence from text with increasing reliability \citep{argyle2023out}; yet this capability varies considerably across model generations and task contexts, and should not be treated as a settled property of LLMs in general. These interpretive questions therefore still require the oversight and judgment of a human researcher, who draws on theoretical knowledge and contextual understanding to validate and refine the model's output. While LLMs can be a valuable tool for identifying potential relationships in a text, they should not be seen as a substitute for the interpretive work of the qualitative analyst \citep{small2011lost}.

\subsection{LLMs for coding, memoing, and theory-building}

One of the most promising applications of LLMs in qualitative SNA is in the area of coding, memoing, and theory-building. These are core practices associated with grounded theory traditions in qualitative research \citep{charmaz2014constructing}. Coding is the process of systematically labeling and organizing qualitative data, while memoing is the process of writing reflective notes about the data and the emerging theoretical ideas. LLMs can be used to assist with both of these tasks. For example, an LLM could be trained to automatically code a large corpus of text based on a predefined coding scheme, or to generate memos based on the patterns it identifies in the data. This can free up the researcher to focus on the more creative and conceptual aspects of theory-building, such as developing new theoretical categories and exploring the relationships between them. However, it is important to remember that LLMs are only as good as the data they are trained on. If the training data is biased, the LLM will reproduce those biases in its analysis. Therefore, it is crucial for researchers to be reflexive about the limitations of their data and to use LLMs in a way that is transparent and accountable \citep{nelson2020computational,abbott2004methods,gilardi2023chatgpt,bail2024can}.

\subsection{LLMs and abductive reasoning}

Abductive reasoning, or inference to the best explanation, is a form of reasoning that is central to the process of theory-building in qualitative research \citep{tavory2014abductive}. It involves moving back and forth between data and theory, generating and refining hypotheses until the most plausible explanation for the data is found. LLMs can be a powerful tool for supporting this process of abductive reasoning by surfacing anomalies in large corpora that would be invisible to a researcher working at a smaller scale.

Consider the following illustration. An LLM analyzing a corpus of organizational communications might identify a cluster of actors who are structurally equivalent -- that is, they occupy the same position in the network, connected to the same alters -- yet whose language, tone, and relational self-descriptions differ markedly. One actor describes their ties in terms of mentorship and obligation; another frames identical structural connections as strategic alliances; a third uses the language of friendship. These divergent relational identities, surfaced by the LLM across hundreds of messages, would be difficult to detect through manual reading alone. The anomaly does not resolve itself automatically: it is the researcher who must then decide whether this variation reflects role conflict, subcultural difference, or a limitation of the network boundary specification. The LLM has done the work of detection; the researcher does the work of interpretation. In this way, LLMs support abductive reasoning not by generating explanations, but by reliably flagging the surprises that make explanation necessary \citep{abbott2004methods,nelson2020computational,tavory2014abductive}.

\subsection{Bias, hallucination, and reflexivity}

Despite their many potential benefits, LLMs also pose a number of significant challenges for qualitative researchers. One of the most serious of these is the problem of bias. LLMs are trained on massive amounts of text from the internet, which is notoriously biased and unrepresentative of the full range of human experience \citep{bender2021dangers}. As a result, LLMs can reproduce and even amplify existing social biases, such as racism, sexism, and classism. Another serious problem is that of ``hallucination,'' in which the LLM generates text that is plausible but factually incorrect. This can be particularly problematic in the context of qualitative research, where the goal is to produce a rich and accurate account of the social world. To mitigate these risks, it is crucial for researchers to be reflexive about the limitations of LLMs and to use them in a way that is transparent and accountable. This means being explicit about the data that was used to train the LLM, the methods that were used to analyze the data, and the potential biases that might be present in the results. It also means subjecting the output of the LLM to careful critical scrutiny, and not taking its pronouncements at face value \citep{argyle2023out}.

\subsection{The risk of semantic homogenization}

When used to scale qualitative interpretation across large corpora, LLMs pose a distinct and under-theorized risk: semantic homogenization. By smoothing over variation in language, context, and meaning, LLM-assisted analysis may inadvertently collapse distinct relational understandings into uniform categories. Crucially, this is not merely a defect of current models that will disappear with more training data; rather, it creates a structural tendency toward semantic homogenization that persists across model generations. Because LLMs operate by predicting the most probable textual continuations based on their training distribution, they structurally favor dominant or majoritarian interpretations of relational cues.

This regularizing effect can obscure minority interpretations, local idioms, or context-specific tie meanings that are central to qualitative network analysis. For instance, a tie that a marginalized sub-community understands as ``protective solidarity'' might be uniformly flattened by the model into a generic ``friendship'' or ``support'' tie, erasing the specific cultural logic that qualitative SNA seeks to uncover. Recognizing and guarding against this homogenizing tendency is essential. Researchers must actively counteract it by prompting models to identify conflicting interpretations, by conducting targeted manual reviews of outlier texts, and by preserving the emic diversity of tie meanings rather than allowing the model to prematurely resolve them into a single etic taxonomy \citep{small2011lost,bender2021dangers}.

\subsection{LLMs as collaborators, not analysts}

LLMs are best positioned as instruments for scaling and systematizing interpretive labor in qualitative and mixed-methods SNA. Rather than producing interpretations autonomously, LLMs can assist researchers by accelerating coding, supporting the exploration of alternative categorizations, identifying relational patterns across large textual corpora, and surfacing cases or anomalies that warrant closer qualitative inspection. In this role, LLMs extend the reach of qualitative analysis while leaving interpretive judgment, theoretical integration, and case-based reasoning firmly in the hands of the researcher \citep{nelson2020computational,abbott2004methods,salganik2018bit}.

It is important to distinguish two different kinds of claims made in this article. The first are \textit{capability claims}: empirically contingent, model-dependent, and time-sensitive assertions about what current LLMs can and cannot do -- claims that are in principle testable and subject to revision as models improve. The second are \textit{methodological commitments}: prior positions about what qualitative research requires, grounded in theoretical and ethical considerations that are independent of any particular model's performance. The recommendation that LLMs serve as collaborators rather than analysts belongs to the second category. It does not rest on the premise that LLMs are incapable of autonomous interpretation; it rests on the premise that interpretive authority, accountability, and reflexivity are constitutive of qualitative inquiry and cannot be delegated to a computational system regardless of its capabilities \citep{maxwell2012qualitative,abbott2004methods}.

\section{Research designs with LLMs + networks}

\subsection{Design patterns for LLM-augmented SNA}

Integrating LLMs into qualitative and mixed-methods SNA requires thoughtful research design. We can identify several key design patterns that offer different ways of leveraging the respective strengths of human researchers and LLMs. These patterns are not mutually exclusive and can be combined in creative ways to suit the specific research questions and data. A primary consideration is the division of labor: what tasks are best suited for the LLM, and what tasks require the nuanced interpretation of a human analyst? For example, LLMs can be used for initial data exploration and pattern detection, while human researchers can focus on the more in-depth analysis of key cases or the development of theoretical insights. The choice of design pattern will also depend on the nature of the data and the goals of the research. Is the goal to generate new theory, to test existing theory, or to produce a rich, descriptive account of a particular social world? By carefully considering these questions, researchers can design studies that are both methodologically rigorous and theoretically innovative \citep{creswell2014mixed}.

In qualitative and mixed-methods network research, this division of labor must be made explicit to prevent the mischaracterization of LLMs as analytic agents rather than methodological supports.

\subsection{Sequential designs: qual \texorpdfstring{$\rightarrow$}{->} quant \texorpdfstring{$\rightarrow$}{->} qual}

One of the most powerful mixed-methods designs is the sequential design, in which qualitative and quantitative methods are used in a phased approach. In the context of LLM-augmented SNA, a common sequential design would be qual $\rightarrow$ LLM $\rightarrow$ qual. In the first phase, the researcher would conduct a small number of in-depth qualitative interviews or ethnographic observations to develop an initial understanding of the social context and to identify the key themes and concepts. In the second phase, the researcher would use an LLM to analyze a much larger corpus of text, using the insights from the first phase to guide the analysis. For example, the researcher could use the LLM to code the text for the themes that were identified in the initial qualitative phase, or to identify all instances of a particular type of social relationship. In the third phase, the researcher would return to the qualitative data, using the findings from the LLM analysis to inform a more focused and in-depth analysis of the key cases. This sequential design allows the researcher to combine the depth of qualitative inquiry with the scale of computational analysis, resulting in a richer and more comprehensive understanding of the social world \citep{creswell2014mixed}.

\subsection{Parallel designs: LLMs + humans}

In a parallel design, the LLM and the human researcher work in tandem, each analyzing the same data but from a different perspective. This allows for systematic triangulation between the computational output of the LLM and the interpretive analysis of the human researcher. For example, the researcher could use an LLM to conduct a topic model of a large corpus of text, while at the same time conducting a close reading of a small sample of the texts. The findings from the two analyses could then be compared and contrasted, with the goal of developing a more comprehensive and nuanced understanding of the data. This parallel design can also be a way to identify the limitations and biases of both the LLM and the human researcher. Where do their analyses converge, and where do they diverge? By exploring these differences, the researcher can gain a deeper understanding of the data and of the research process itself \citep{small2011lost}.

Crucially, this parallel design requires a specified procedure for validation and disagreement analysis. When an LLM is used to code relational data or classify ties, researchers should not simply accept the output. Instead, they must establish a human-machine intercoder reliability protocol. This involves having human researchers manually code a representative subset of the corpus and systematically comparing these codes against the LLM's output using standard reliability metrics (e.g., Cohen's kappa or Krippendorff's alpha). However, unlike traditional intercoder reliability where the goal is mere agreement, divergence in the human-machine case should be treated as an analytic opportunity. Disagreement analysis -- where researchers closely inspect the instances where human and LLM codes diverge -- often reveals ambiguities in the coding scheme, nuances in local idioms that the model missed, or unexamined assumptions on the part of the human coder. The human researcher retains ultimate interpretive authority to correct the LLM's output, but the systematic comparison provides a rigorous foundation for those corrections. As a practical decision rule, we recommend that if agreement falls below a threshold of $\kappa = 0.70$, the researcher should revise the prompt before proceeding to full-corpus coding; if disagreement exceeds 20\% on the validation sample, a full recoding pass with a revised scheme is warranted. These thresholds are illustrative rather than universal: acceptable agreement levels depend on the complexity of the coding scheme, the nature of the research question, and the consequences of misclassification in the specific study \citep{nelson2020computational}.

\subsection{Validity, transparency, and audit trails}

As with any research method, the use of LLMs in qualitative SNA raises important questions of validity, transparency, and accountability. To ensure that LLM-augmented analyses remain transparent and auditable, researchers must develop clear and systematic documentation procedures. While qualitative research has long relied on audit trails to document analytic decisions, LLM-augmented research introduces novel, system-specific barriers to reproducibility. These include the inherent nondeterminism of language models across runs (even at temperature zero), silent API updates by providers, and the total loss of reproducibility when a provider retires a specific model version. A robust audit trail for LLM-augmented SNA must therefore go beyond generic documentation. It must explicitly record the exact model version used (e.g., \texttt{gpt-4-0613} rather than just ``GPT-4''), the precise prompts deployed, the hyperparameter settings (such as temperature and top-p), and the date of access. Furthermore, because API-based models can be deprecated, researchers should archive the raw, unedited LLM outputs alongside the final, human-corrected data. By adhering to these LLM-specific transparency standards, researchers can ensure that their mixed-methods SNA remains methodologically rigorous and auditable even as the underlying technologies change \citep{creswell2014mixed,small2011lost}.

\section{Transition to practice}

\subsection{Software environments and workflows}

For computational social scientists, the Jupyter notebook has become a standard environment for LLM-augmented SNA, allowing researchers to script API calls, process data with libraries like \texttt{pandas}, and visualize networks using \texttt{networkx} \citep{salganik2018bit}. However, we must emphasize that adopting LLMs for qualitative SNA does not require qualitative researchers to become Python programmers. The transition to practice is already being facilitated by the integration of LLM capabilities directly into the graphical user interfaces (GUIs) of established Computer-Assisted Qualitative Data Analysis Software (CAQDAS), such as MAXQDA, NVivo, and ATLAS.ti, as well as dedicated network analysis tools like Gephi or UCINET \citep{borgatti2009network}. A typical qualitative workflow might involve importing interview transcripts into a CAQDAS environment, using built-in LLM tools to generate provisional relational codes or summarize ties, manually reviewing and correcting those codes within the software's interface, and then exporting the validated edge list to a network visualization tool. Whether executed via Python scripts or GUI-based software, the core methodological requirement remains the same: the environment must allow the researcher to inspect the raw text, the LLM's prompt, and the resulting output side by side, enabling the seamless correction of computational inferences before they are formalized into network structures.

\subsection{Data ethics, consent, and responsible use}

The use of LLMs in qualitative SNA raises a number of important ethical considerations. One of the most pressing is the issue of data privacy and informed consent. When we collect data from online sources, we are often dealing with the personal data of individuals who have not explicitly consented to be part of our research. While it may be legal to collect this data, it is not always ethical. Researchers have a responsibility to protect the privacy of their research subjects and to ensure that their data is not used in ways that could harm them. This is particularly important when working with vulnerable populations or with sensitive topics. Another ethical consideration is the potential for bias and discrimination \citep{weidinger2021ethical}. As we have already discussed, LLMs can reproduce and even amplify existing social biases. This can have serious consequences, particularly if the research is used to inform policy or to make decisions about individuals. To mitigate these risks, it is crucial for researchers to be transparent about their methods and their data, and to take steps to ensure that their research is conducted in a way that is fair and equitable. This includes being mindful of the potential for algorithmic bias, and taking steps to de-bias their models and their data \citep{salganik2018bit,small2011lost}.

\subsection{An illustrative worked example}

To ground the methodological framework developed in this article, we sketch a minimal illustrative worked example that illustrates the full LLM-augmented SNA pipeline. Consider a corpus of 200 semi-structured interview excerpts in which participants describe their professional relationships within an organization. The researcher begins by formulating a structured prompt instructing the LLM (e.g., GPT-4, model version \texttt{gpt-4-0613}, temperature = 0) to identify all directed relational ties mentioned in each excerpt and to classify each tie along two dimensions: its \textit{type} (e.g., mentorship, collaboration, and rivalry) and its \textit{valence} (positive, negative, and ambivalent).

The raw LLM output for a representative excerpt might read: \textit{``Tie: Ego $\rightarrow$ Alter A; Type: mentorship; Valence: positive. Tie: Ego $\rightarrow$ Alter B; Type: rivalry; Valence: negative.''} The researcher then inspects a stratified random sample of 20\% of all outputs, correcting misclassifications -- for instance, where the model conflated a tie of ``collegial obligation'' with ``friendship'' -- and documenting each correction in a separate audit log. The corrected edgelist is then imported into a network analysis environment (e.g., \texttt{networkx} in Python or Gephi) to construct a directed, signed network. Structural analysis of this network -- including measures of reciprocity, clustering, and positional equivalence -- is then interpreted in light of the qualitative tie meanings preserved from the original coding.

This example is intentionally schematic; its purpose is not to prescribe a single workflow but to demonstrate that the five components identified by the reviewer -- corpus, prompt, raw LLM output, researcher corrections, and resulting network -- can be made explicit and transparent within a single, auditable research process. Supporting materials, including prompt templates and annotated workflow scripts, are available at \url{https://github.com/mboudour/ai-augmented-research}.

\section{Conclusions and further development}

In this manuscript, we have argued that the integration of LLMs into qualitative and mixed-methods SNA has the potential to be a transformative development for the field. By automating some of the more laborious aspects of qualitative data analysis, LLMs can allow researchers to analyze larger datasets, identify more subtle patterns, and generate new theoretical insights. When used carefully, LLMs make it possible to extend qualitative and mixed-methods SNA to empirical settings that were previously inaccessible due to scale. They enable researchers to trace relational meanings across large volumes of communication, to compare interpretive patterns across cases, and to integrate qualitative insight more systematically with formal network analysis. The contribution of LLMs, therefore, lies not in replacing qualitative inquiry, but in expanding its scope, reach, and connective capacity within network research \citep{salganik2018bit,nelson2020computational}.

However, we have also cautioned against a naive or uncritical adoption of these new technologies. LLMs are not a panacea for the challenges of qualitative research. They are powerful tools, but they are also limited and flawed. They can be biased, they can hallucinate, and they can be used in ways that are unethical and irresponsible. Therefore, it is crucial for researchers to approach the use of LLMs with a critical and reflexive mindset. We must be mindful of the limitations of these technologies, and we must take steps to ensure that we are using them in a way that is methodologically rigorous, ethically sound, and theoretically informed. The future of qualitative SNA will not be one in which human researchers are replaced by machines. Rather, it will be one in which human researchers and machines work together in a collaborative partnership, each leveraging their respective strengths to produce a richer and more comprehensive understanding of the social world. The challenge for our field is to develop the methods, the theories, and the ethical frameworks that will allow us to realize the full potential of this exciting new frontier.

\section*{Acknowledgements}
I am grateful to E. and G. for their emotional support.

\section*{Author contributions}
Moses Boudourides was responsible for the concept, writing, and review of the article.

\section*{Funding information}
The author has no funding to disclose.

\section*{Conflict of interest}
The author declares no conflict of interest.

\bibliographystyle{plainnat}
\bibliography{references}

\end{document}